\documentclass[aps,prl,epsfigure,twocolumn,showpacs,superscriptaddress]{revtex4-1}
\usepackage{color}

\usepackage{amsmath}
\usepackage{amssymb,amsthm}
\usepackage{verbatim}
\usepackage{epsfig}
\usepackage{epstopdf}

\begin{document}

\title{Entangled states in the role of witnesses
}

\author{Bang-Hai~Wang
\footnote{Correspondence to:
bhwang@gdut.edu.cn}}
\affiliation{School of Computer Science and Technology, Guangdong University of Technology, Guangzhou 510006, China}
\affiliation{Centre for Quantum Technologies, National University of Singapore, 3 Science Drive 2, Singapore 117543, Singapore}
\affiliation{Department of Computer Science, and Department of Physics and Astronomy, University College London, Gower Street, WC1E 6BT London, United Kingdom
}%
\date{\today}

\begin{abstract}
Quantum entanglement lies at the heart of quantum mechanics and quantum information processing. In this work, we show a new framework where entangled states play the role of witnesses. We extend the notion of entanglement witnesses, developing a hierarchy of witnesses for classes of observables. This hierarchy captures the fact that entangled states act as witnesses for detecting entanglement witnesses and separable states act as witnesses for the set of non-block-positive Hermitian operators. Indeed, more hierarchies of witnesses exist. We introduce the concept of \emph{finer} and \emph{optimal} entangled states. These definitions not only give an unambiguous and non-numeric quantification of entanglement and a new perspective on edge states but also answer the open question of what the remainder of the best separable approximation of a density matrix.
Furthermore, we classify all entangled states into disjoint families with optimal entangled states at its heart. This implies that we can focus only on the study of a typical family with optimal entangled states at its core when we investigate entangled states. Our framework also assembles many seemingly different findings with simple arguments that do not require lengthy calculations.

\end{abstract}


\maketitle

\noindent \textit{Introduction.---} Quantum correlations, especially quantum entanglement have been recognized as a novel resource that may be used for tasks that are either very inefficient or impossible in the classical realm \cite{Horodecki09,Guhne09,Plenio07}. However, quantum entanglement has not been fully understood. An effective method has not yet been found to detect whether or not a given state is entangled. And even if a given mixed state is known to be entangled, quantifying the amount of entanglement it contains is hard. In this work, we show \emph{entanglement witnesses} can unequivocally answer both questions. 

Another essential approach in the study of entanglement comes from the best separable approximation (BSA) decomposition \cite{Lewenstein98} (also called Lewenstein-Sanpera decomposition) of a density matrix. The BSA of an arbitrary state $\rho$ was defined from its convex decomposition as $\rho=\lambda\rho^s+(1-\lambda)\rho^{E}$, where $\rho^s$ is a separable state, $\rho^{E}$ is a state that does not have any product vector in its range, and the real parameter $\lambda$ is maximal. The separable state $\rho^s$ is called the best separable approximation of $\rho$. The Lewenstein-Sanpera decomposition was based on subtracting projections on product vectors from a given density matrix in such a way that \emph{the unique remainder} remains positive semi-definite. This approach can naturally serve as a quantification of entanglement and allowed for the derivation of many very strong results \cite{Sanpera98,Kraus00,Horodecki00,Lewenstein00,Karnas01}. Various works have developed on this topic \cite{Thiang09,Thiang10,Quesada14,Lewenstein16}, but the remainder of the Lewenstein-Sanpera decomposition has remained unknown \cite{Karnas01}.  Furthermore, how to parametrize the remainders (the so-called edge states in the case of positive partial transposition entangled states) still remains open \cite{Karnas01,Chruscinski14}.

In this work, we fill in this gap by introducing the hierarchies of witnesses. Entanglement witnesses, entangled states, separable states, and so on constitute a hierarchy of witnesses, each detecting a different class of operator. We then define the notions of finer and optimal entangled states. Theses definitions show an unambiguous and non-numeric quantification of entanglement. We show that the optimal entangled state corresponds to the remainder of the Lewenstein-Sanpera decomposition and the edge state is typical of the optimal entangled state. We further unambiguously classify all entangled states into disjoint families each with a single optimal entangled state at its core. Finally, we show some known finds with simple arguments.

\noindent \textit{The hierarchies of ``witnesses".---}
A remarkable research effort has been devoted to detecting and quantifying entanglement \cite{Guhne09,Horodecki09}. The method of \emph{entanglement witnesses} is currently considered to be the most important and best method for detecting entanglement \cite{Augusiak11}. It is known that the set of separable states is convex and compact. For any entangled state, by the Hahn-Banach theorem \cite{Edwards65} there exists at least one operator that can be used to detect it. Such operators were investigated in the field of the quantum theory because the corresponding positive maps were rediscovered by Peres and Horodecki \cite{Horodecki09, M.Horodecki96}, and later they were called entanglement witnesses by Terhal \cite{Terhal00} - stressing their physical importance as entanglement detectors.
More precisely, an entanglement witness is a Hermitian operator, $W=W^\dag$, such that
(i) $\text{tr}(W\sigma)\geq0$ for all separable states $\sigma$, and
(ii) there exists an entangled state $\pi$ such that $\text{tr}(W\pi)<0$.
Entanglement witnesses have raised considerable attention \cite{Lewenstein98,Doherty04,Wu06,Wu07,Sperling09,Chruscinski09,Hou10a,Chruscinski10,Wang11} (for a recent review, see Ref. \cite{Chruscinski14}). Unfortunately, constructing them for a given entangled state is a difficult task, and the determination of entanglement witnesses for all entangled states is a nondeterministic polynomial-time (NP) hard problem \cite{Gurvits04,Doherty04,Hou10a}.

Entanglement witnesses were introduced because we cannot directly detect entanglement. Constructing entanglement witnesses in general, and finding the minimal set of them that allows for the detection of all entangled states is one of the most challenging open questions \cite{Lewenstein01}. When an entanglement witness $W$ detects an entangled state $\pi$, we say that $W$ ``witnesses" the entanglement of $\pi$. How do we tell if an operator is an entanglement witness? Put another way, what ``witness'' entanglement witnesses?  We denote by $\mathbf{S}$ the set of all separable states, $\mathbf{E}$ the set of all entangled states, $\mathbf{W}$ the set of all entanglement witnesses, and $\mathbf{Q}\equiv\mathbf{S}\cup\mathbf{E}$ the set of all quantum states. Figure 1(a) illustrates the schematic picture.

\begin{figure}[htbp]
\epsfig{file=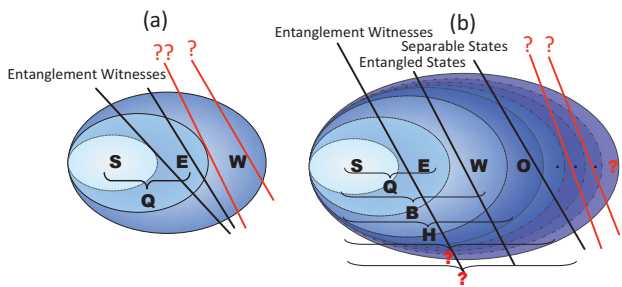,width=.95\columnwidth}
\caption{(Color online) (a) What ``witness" entanglement witnesses? (b) A hierarchy structure of ``witnesses".  }
\label{fig1}
\end{figure}

It is known that the set of quantum states (separable states and entangled states) is also convex and compact. Hence, by the Hahn-Banach theorem, there is at least one ``super" or in other word ``high-level" witness ``witnessing" an entanglement witness \cite{Brub2002}.

For a high-level witness  $\Pi$,

(i') ${\rm tr}(\Pi\rho)\geq0$ for all quantum state $\rho$ (entangled or not);

(ii') there exists an entanglement witness $W$ such that ${\rm tr}(\Pi W)<0$.

Operators that satisfy the above two conditions and play the role of high-level witnesses are none other than entangled states. Entanglement witnesses ``witness'' entangled states and entangled states ``witness'' entanglement witnesses.
While this answer is not difficult to obtain, the role of entangled states as high-level witnesses motivates the study of entangled states in a new way. It follows that the quantum states, entanglement witnesses, and so on are both operators on the Hilbert space and also points in a closed convex set in a real Hilbert space. It is this dual role which underlies our analysis.


Furthermore, one may ask whether or not there exist ``\emph{higher}-level" witnesses that separate other observables from the set of the quantum states (separable and entangled) and entanglement witnesses, as shown in Fig. 1 (b). To investigate these, let us consider the set of bounded Hermitian operators, which have positive expectation values for separable states
\begin{equation}
\mathbf{\mathbf{B}}\equiv\left\{b\,|b=b^\dag, \textrm{tr}\left(\sigma b\right)\geq 0 \quad\forall \sigma\in\mathbf{S}\right\}.
\end{equation}
The set $\mathbf{B}$ is called the set of block-positive \cite{Chruscinski14, Milne15} or partial positive
operators \cite{Wu06,Sperling09}. In standard quantum mechanics, all observables are mathematically denoted by Hermitian operators. We can also separate other observables from entanglement witnesses.
We can easily conclude that $\mathbf{B}=\mathbf{S}\cup\mathbf{E}\cup\mathbf{W}$ and $\mathbf{B}$ is also convex and compact.

To investigate these ``higher-level" witnesses, let us consider the set of bounded Hermitian operators
\begin{equation}
\mathbf{O}\equiv\mathbf{H}-\mathbf{S}-\mathbf{E}-\mathbf{W},
\end{equation}
where $\mathbf{H}=\{h|h=h^\dag\}$ denotes the set of all Hermitian operators.
For a ``higher-level" witness $\Xi$, (i'') ${\rm tr}(\Xi b)\geq0$ for an arbitrary block-positive operators $b\in \mathbf{B}$ (a quantum state or an entanglement witness); (ii'') there exists a non-block-positive observable $o\in\mathbf{O}$ such that ${\rm tr}(\Xi o)<0$.
We can conclude that the ``higher-level" witnesses are just the separable states. Separable states separate entanglement witnesses from the non-positive-and-non-entanglement-witness observables. Sometimes, the measurement of non-Hermitian operators \cite{Bender98,Moiseyev11} occurs in quantum mechanics \cite{Pati15}. We can also find the witnesses of non-Hermitian operators. Mathematically, we can construct more and more convex and compact sets such that they include the set of Hermitian operators. This leads to a hierarchy of ``witnesses", and a question as to whether or not there are ``infinite higher-levels" of witnesses, physically, or mathematically, as shown in Fig. 1 (b).

If an entanglement witness can be written in the form
\begin{equation}\label{DecomposableEW}
W_d=aP+(1-a)Q^\Gamma,
\end{equation}
where $a\in[0,1]$, $P\geq0$, and $Q\geq0$, the entanglement witness is called decomposable \cite{Lewenstein00}. If it does not admit this form, it is called non-decomposable. The set of decomposable entanglement witnesses $\mathbf{W_d}$ is convex and compact \cite{Lewenstein00}.
There exist ``witnesses" separating non-decomposable entanglement witnesses from decomposable entanglement witnesses. Clearly, these ``witnesses" are just bound entangled states. Moreover, all quantum states can be written in the form of Eq. (\ref{DecomposableEW}) and the set $\mathbf{D}\equiv\mathbf{S}\cup\mathbf{E}\cup\mathbf{W_d}$ is also convex and compact \cite{Sarbicki07}. There exist ``witnesses" detecting the observables out of the set $\mathbf{D}$ (here we call $\mathbf{D}$ the set of decomposable observables). Therefore, the decomposable observables, the bound entangled states, and the separable states form another different hierarchy of ``witnesses". Recently, the concept of the coherence witness was put forward and the relation was revealed between the coherence witness and the robustness of coherence \cite{Piani16,Napoli16}. The coherence witnesses, the coherent states (as high-level coherence witnesses \cite{SuperCW}), and the Non-Hermitian witnesses form another hierarchy of ``witnesses". We denote $\mathbf{I}$ the set of incoherent states, $\mathbf{C}$ the set of coherent states, and $\mathbf{W_n}$ the set of non-decomposable entanglement witnesses.  Figure 2(a) illustrates the schematic picture.

\begin{figure}[htbp]
\epsfig{file=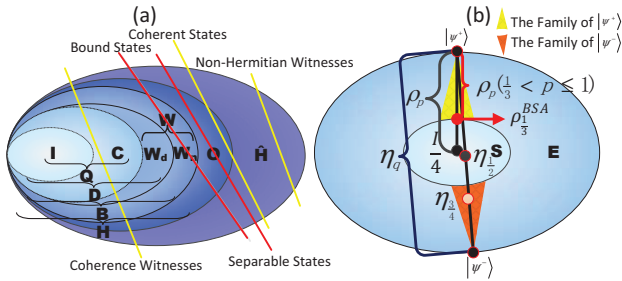,width=.95\columnwidth}
\caption{(Color online) (a) More hierarchies of ``witnesses". (b) Two families (the Werner states as a part of members) of entangled states in $\mathbb{C}^2\otimes\mathbb{C}^2$}
\label{fig2}
\end{figure}

Any quantum state can be mathematically considered as a tight and convex set. There exist hyperplanes between any two different states by the Hahn-Banach theorem. Therefore, there exist many ``witnesses" which are ubiquitous in nature. All in all, we can find and even mathematically construct more witnesses (e.g. ultrafine entanglement witnessing \cite{Shahandeh17}) and more hierarchies of witnesses (e.g. Schmidt-number witnesses \cite{Terhal2000,Sanpera01}).

Once we find that there exist ``witnesses" between two sets, how do we generally construct them? If $S_1$, $S_2$ are convex closed sets in a real Banach space and one of them is compact, there exists a continuous functional $f$ and $c\in \mathbb{C}$ such that for all pairs $e_1\in S_1$, $e_2\in S_2$, we have $f(e_1)<c\leq f(e_2)$ \cite{M.Horodecki96}. It is known that any continuous functional $f$ on a Hilbert space can be represented by a vector from this space. Any linear functional $g$, acting on trace class operators $\rho$, can be written as $g(\rho)={\rm tr}(\rho H)$ for a bounded, Hermitian operator $H$ \cite{Bruns16}. Therefore, we need to find Hermitian operators, which ``witness" $\rho_2\in S_2$, such that $\min\{{\rm tr}(e_2 H)\}<{\rm tr}(\rho_2 H)\leq \min\{{\rm tr}(e_1 H)\}$. Moreover, the optimization problem (under certain constraints) can be solved by the method of Lagrange's multipliers \cite{Sperling09,Bruns16}.

\textit{The entangled states as high-level witnesses.---}
Inspired by the investigation for entanglement witnesses \cite{Lewenstein00}, we define as follows. Given a high-level witness (entangled state) $\rho$, define $D_\rho=\{W|{\rm tr}(\rho W)<0\}$; that is the set of operators ``witnessed" by $\rho$. Given two high-level witnesses, $\rho_1$ and $\rho_2$, we say that $\rho_2$ is finer than $\rho_1$, if $D_{\rho_1}\subseteq D_{\rho_2}$; that is, if all the operators ``witnessed" by $\rho_1$, are also ``witnessed" by $\rho_2$. We say that $\rho$ is an optimal high-level witness if there exists no other high-level witness which is finer.

Naturally, we have the properties and characterization of entangled states as high-level witnesses.

\textbf{ Lemma 1.} (\cite{Wu07}) Let $\rho_2$ be finer than $\rho_1$ and $\delta\equiv \inf_{W_1\in D_{\rho_1}} |\frac{{\rm tr}(W_1\rho_2)}{{\rm tr}(W_1\rho_1)} |$.
Then we have the following:
(i) If ${\rm tr}(W\rho_1)=0$, then ${\rm tr}(W\rho_2)\leq 0$;
(ii) If ${\rm tr}(W\rho_1)<0$, then ${\rm tr}(W\rho_2) \leq {\rm tr}(W\rho_1)$;
(iii) If ${\rm tr}(W\rho_1)>0$, then $\delta {\rm tr}(W\rho_1)\geq {\rm tr}(W\rho_2)$;
(iv) $\delta\geq 1$. In particular,
$\delta=1$ iff $\rho_1=\rho_2$.

\textbf{ Corollary 1.} (\cite{Wu07}) $D_{\rho_1}=D_{\rho_2}$ if and only if $\rho_1=\rho_2$.

This result tells us that we can completely characterize entangled states by entanglement witnesses.

If we replace ``finer" with ``more entangled" in the previous definition, we can immediately get a characterization of how entangled a given state is. Given two entangled states, $\rho_1$ and $\rho_2$, we say that $\rho_2$ is more entangled than $\rho_1$, if $D_{\rho_1}\subseteq D_{\rho_2}$. We say that $\rho$ is an optimal entangled state if there exists no other high-level witness which is more entangled. This definition shows the ``witnessing power" of entangled states.
Let us express the above idea in a more rigorous way.

\textbf{Theorem 1.} All entangled states are unambiguously quantified by the sets of observables that are ``high-level-witnessed" by the entangled state.

It is easy to conclude that this entanglement measure is a ``good" measure of entanglement by the axiomatic approach to quantifying entanglement \cite{Vedral97}: (i) all sets of entanglement witnesses for separable states are empty; (ii) local unitary operations leave it invariant; and (iii) this measure of entanglement cannot increase under Local Operations and Classical Communication (LOCC) \cite{Plenio07}.
Alternatively, the entanglement content of a state can be quantified via $E(\rho)=\max\{0,-\min_{W\in M}{\rm tr}(W\rho) \}$,
where $M$ is the intersection of the set of entanglement witnesses with some other set $C$, such that $M$ is compact \cite{Brandao05}. In contrast, our entanglement measure is quantified by the number of ``witnesses" that are ``witnessed" by the entangled states, not by a numerical value.

Generally, there are two categories of entanglement measures \cite{Horodecki09}. One is based on definitions of operational tasks. This entanglement measure has a directly physical implication, but it is generally difficult to compute. A variety of such entanglement measures (for example, entanglement cost and entanglement of distillation) are NP-hard to compute \cite{Huang14}. The second category is more axiomatic \cite{Vedral97}. An entanglement measure (also called an entanglement monotone) is often an operator function satisfying several basic properties. Examples include entanglement of formation \cite{Bennett96}, concurrence \cite{Hill97,Wootters98}, negativity \cite{Zyczkowski98,Vidal02}, Schmidt number \cite{Shahandeh14,Terhal2000}, and so on \cite{Regula16} (for a review, see Ref.\cite{Plenio07}). Almost all known criteria, however, map an entangled state to a real number (sometimes between 0 and 1 for comparison). Generally whether a state is regarded as being more entangled than another is dependent on the choice of criterion used. Therefore, there exists the possibility that two different entangled states obtain the same value with respect to some measure. There also exists the possibility that a criterion indicates $\rho_1$ is more entangled than $\rho_2$ but another criterion shows $\rho_2$ is more entangled than $\rho_1$.

Here we have quantified the entanglement of a state by the entanglement witnesses that are ``high-level-witnessed'' by the state. For two different entangled states, there exist different sets of ``witnesses'' that characterize them. This result indicated that two different, optimal entangled states are incomparable -- one can not say that one is more entangled than the other just as we cannot say which one is ``finer" between the badminton and tennis world champions.

\textit{The remainder of the best separable approximation decomposition and the structure of entangled states.---} Furthermore, we have the following properties of entangled states in the role of witnesses.

\textbf{ Lemma 2.} $\rho_2$ is finer (more entangled) than $\rho_1$ if and only if there exists
$1>\epsilon\geq 0$ such that $\rho_1=(1-\epsilon)\rho_2+\epsilon P$, where
$P$ is not finer than $\rho_1$ or it is separable.

{\it Proof.---} (If) For all $W\in D_{\rho_1}$ we have that
$0>{\rm tr}(W\rho_1)= (1-\epsilon){\rm tr}(W\rho_2)+\epsilon {\rm tr}(WP)$
which implies ${\rm tr}(W\rho_2)<0$ and therefore $W\in D_{\rho_2}$. (Only
if) We define $\delta$ as in Lemma 1. Using Lemma 1(iv) we have
$\delta\ge 1$. First, if $\delta=1$ then according to Lemma 1(iv) we
have $\rho_1=\rho_2$ (i.e., $\epsilon=0$). For $\delta> 1$, we define
$P=(\delta-1)^{-1}( \delta \rho_1-\rho_2)$ and $\epsilon=1-1/\delta>0$.
We have that $\rho_1=(1-\epsilon)\rho_2+\epsilon P$, so that it only remains
to be shown that $P\ge 0$. But this follows from Lemma 1(i--iii) and the
definition of $\delta$, $\delta=\inf_{W_1\in D_{\rho_1}} \left|
\frac{{\rm tr}(W_1\rho_2)}{{\rm tr}(W_1\rho_1)} \right|$. We can easily know $P$ is not finer than $\rho_1$ or it is separable. $\Box$

\textbf{ Corollary 2.} $\rho$ is optimal if and only if for all projectors on product vectors $P$ and
$\epsilon>0$, $\rho'=(1+\epsilon)\rho-\epsilon P$ is not a high-level witness (legitimate entangled state).

{\it Proof.---} (If) According to Lemma 2, there is no entangled state which is
finer than $\rho$, and therefore $\rho$ is optimal. (Only if) If $\rho'$
is an entangled state, then according to Lemma 2 $\rho$ is not optimal. $\Box$


By Corollary 2, we can conclude that we can construct optimal entangled states by the technique of ``subtracting projectors on product vectors" \cite{Lewenstein00}. We have the following result.

\textbf{Theorem 2.} An arbitrary (normalized) density matrix $\rho$ has a \emph{unique} decomposition in the form
\begin{equation}
 \rho=\Lambda\rho^{S}+(1-\Lambda)\rho^{optE}; \Lambda\in [0,1],
\end{equation}
where $\rho^{optE}$ denotes the optimal entangled state of $\rho$, $\rho^{S}$ denotes the BSA \cite{Lewenstein98,Karnas01} of the density matrix, and $\Lambda$ is maximal.

What is the remainder of the Lewenstein-Sanpera decomposition is yet to be known \cite{Lewenstein98,Karnas01}. It is not difficult to conclude that the procedure of optimization for a general entangled state $\rho$ is merely to find the Lewenstein-Sanpera decomposition of $\rho$, and the remainder of the Lewenstein-Sanpera decomposition of $\rho$ is just the optimal entangled states.
If we subtract any projector onto a product vector from a positive partial transposition entangled state (PPTES), then the resulting operator is no longer a PPTES. It is called an edge state \cite{Lewenstein00,Kraus00,Horodecki00,Lewenstein98,Sanpera98,Leinaas07,Lewenstein01}, because it lies on the edge between PPTES's and entangled states with non-positive partial transposition.
However, the complete characterization of edge states is lacking in the literature \cite{Chruscinski14}. Our results show edge states are optimal entangled states and they can be generalized to the so-called $k$-edge state \cite{Sanpera01}.

According to our results and the range criterion \cite{Horodecki97}, the definitions of completely entangled subspace (CES) \cite{Wallach02,Parthasarathy04,Cubitt08,Walgate08}, we can easily conclude the following results.

\textbf{Remark 1.} If a quantum state is such that its range does not contain any product vector $|e,f\rangle$, then it is an optimal entangled state.



\textbf{Remark 2.} If the support of an entangled state $\pi$ does not contain any product vector or $\textrm{Support}(\pi)$ is a CES, which does not contain any product state, then $\pi$ is an optimal entangled state.

\textbf{Remark 3.} All mixed states on a CES are optimal entangled states.

We need the following results before we sketch the proof of Theorem 2. Note that the uniqueness of the Lewenstein-Sanpera decomposition in any bipartite system was also proven by Karnas and Lewenstein in a different way in Ref. \cite{Karnas01}.

\textbf{Lemma 3.} For an entanglement witness $W$, there exist one and only one optimal entangled state $\rho^{optE}$ to ``high-level-witness" it.

{\it Proof.---} It is clear there exists at least one optimal entangled state to ``high-level-witness" it for an entanglement witness. Suppose two different optimal entangled states $\rho^{optE}$ and $\rho^{optE'}$ ``high-level-witness" the same entanglement witness $W$. By Corollary 4 (below in the main text, also see \cite{Wu07}), $\rho_p=p\rho^{optE}+(1-p)\rho^{optE'}$ is a high-level witness (entangled state) for $0\leq p\leq 1$, and we can find a sufficiently small $p^*$ such that $W$ high-level-witnessed by $\rho_{p^*}$ (i.e., ${\rm tr}(W\rho_{p^*})<0$) since ${\rm tr}(W\rho^{optE})<0$ is bound. Thus, $D_\rho\supseteq D_{\rho^{optE'}}$. By the optimality of $\rho^{optE'}$, $D_{\rho}=D_{\rho^{optE'}}$. In the same way, $D_{\rho}\supseteq D_{\rho^{optE}}$ and $D_{\rho}=D_{\rho^{optE}}$. By Corollary 1, $\rho^{optE}=\rho^{optE'}$.
$\Box$

\textbf{ Corollary 3.} For an arbitrary entangled state, there exists one and only one optimal entangled state.

\textbf{ Lemma 4.} For any density matrix $\rho$ (separable, or not) and for any set $V$ of product vectors belonging to the range of $\rho$, i.e., $|e,f\rangle \in \mathcal{R}(\rho)$, there exist a separable (in general not normalized) matrix
\begin{equation}
\rho^{S}=\sum_\alpha\Lambda_\alpha P_\alpha
\end{equation}
with all $\Lambda_\alpha\geq0$, such that $\rho^{optE}=\rho-\rho^{S}\geq0$, and $\rho^{S}$ provides the \emph{unique} BSA to $\rho$ and $\rho^{optE}$ provides the (unnormalized) optimal entangled state to $\rho$ in the sense that the trace ${\rm tr}(\rho^{optE})$ is minimal (or, equivalently, ${\rm tr}\rho^{S}\leq1$ is maximal).

{\it Proof.---} By Corollary 2 and Ref. \cite{Lewenstein98}, we can know $\rho^{optE}$ is an optimal entangled state while ${\rm tr}\rho^{S}\leq1$ is maximal. Different from the result in Ref. \cite{Lewenstein98}, now we only need to show the uniqueness of the BSA to $\rho$. Following from Ref. \cite{Lewenstein98}, the trace of $\rho^{S}$ is unique, and the trace of $\rho^{optE}$ is unique. By Corollary 3, normalized $\rho^{optE}$ is unique. Given $\rho$, $\rho^{S}=\rho-\rho^{optE}$, the BSA of $\rho$ is also unique.
$\Box$

Now we show our proof of Theorem 2.

{\it Proof.---} By normalizing the optimal entangled state $\rho^{optE}$ and the BSA $\rho^{S}$ of $\rho$ in Lemma 4, we can easily conclude Theorem 2.
$\Box$

As an immediate consequence, we obtain an unambiguous classification of entangled states.

\textbf{Theorem 3.} The set of entangled states is composed of disjoint families. Each family contains a single optimal entangled state and the other members of the family are obtained by mixing this optimal entangled state with product states.

This result implies a family structure of entangled states. Our results indicate that we can restrict ourselves to the study of a typical family centered around an optimal entangled state when we investigate entangled states.

\textit{Different findings with simple arguments.---}
Our framework assembles many seemingly different findings with simple arguments.
Let us first focus on the question of when different entanglement witnesses can detect the same entangled states.

\textbf{Lemma 5.} \cite{Wu06} There exists an entangled state $\rho$ detected by $W_1$ and $W_2$ if and only if for all $\lambda\in [0,1]$, $W=\lambda W_1+(1-\lambda)W_2$ is not a positive operator (in other words, $W=\lambda W_1+(1-\lambda)W_2$ must be an entanglement witness because ${\rm tr}(W_1\sigma)\geq0$, ${\rm tr}(W_2\sigma)\geq0$ implies that ${\rm tr}(W\sigma)\geq0$ for all separable states $\sigma$).

Since entangled states are also (high-level) witnesses, this question can be changed into the question of when different high-level witnesses (entangled states) can detect the same entanglement witness.

\textbf{Corollary 4.} There exists an entanglement witness $W$ detected by a high-level witness (entangled state) $\Pi_1$ and a high-level witness (entangled state) $\Pi_2$ if and only if for any $\lambda\in [0,1]$, $\Pi=\lambda \Pi_1+(1-\lambda)\Pi_2$ is a high-level witness (entangled state).

This recovers the main result of Ref.\cite{Wu07}.

Since the maximum dimension of the CES subspace in $\mathcal{H}=\mathcal{H}_1\otimes\mathcal{H}_2\otimes\cdots\otimes\mathcal{H}_k$ is $d_1d_2\cdots d_k-(d_1+d_2+\cdots\cdots+d_k)+k-1$ ($d_i$ denotes the dimension of $\mathcal{H}_i$ ) \cite{Parthasarathy04}, we can immediately conclude the following result.

\textbf{Corollary 5.} A high-level witness in $\mathcal{C}^2\otimes\mathcal{C}^2$ is optimal if and only if it is a pure entangled state.

This recovers one of the main results in Ref.\cite{Lewenstein98}.

To illustrate these concepts, consider the Werner state
\begin{equation}
\rho_p=p|\psi^+\rangle\langle\psi^+|+(1-p)\frac{I}{4},
\end{equation}
where $|\psi^+\rangle=\frac{1}{\sqrt{2}}\left(|00\rangle+|11\rangle\right)$ and $0\leq p\leq1$~\cite{Werner89}. It is known that $\rho_p$ is entangled for $\frac{1}{3}<p\leq1$. Therefore, $\rho_p$ is  a high-level witness for $\frac{1}{3}<p\leq1$, and the set ``high-level-witnessed" by $\rho_p$, $D_{\rho_p}=\{W_p|W_p\ngeq0, \textrm{tr}\left(\rho_pW_p\right)<0 \}$.



 We can easily determine that $W_p$ takes the form  $W_p=q|\varphi\rangle\langle\varphi|^\Gamma+(1-q)\varrho$ such that $\textrm{tr}\left(\rho_pW_p\right)<0$, where $0\leq q\leq1$, $|\varphi\rangle=\frac{1}{\sqrt{2}}(|10\rangle-|01\rangle)$, and $\varrho$ is a quantum state. This follows from the fact that the entanglement witness $W=|\varphi\rangle\langle\varphi|^\Gamma$ is optimal for this state \cite{Augusiak11} and the eigenvector of the negative eigenvalue of $W$ is just $|\psi^+\rangle$.
 Note that $\rho_s$ is finer (more entangled) than $\rho_t$ for $\frac{1}{3}<s<t\leq1$, and $|\psi^+\rangle$ is the optimal entangled state because the Lewenstein-Sanpera decomposition of $\rho_p$ is
 \begin{equation}
 \rho_p=\frac{1}{2}(3-3p)\rho_{\frac{1}{3}}^{BSA}+\frac{1}{2}(3p-1)|\psi^+\rangle\langle\psi^+|,
 \end{equation}
 for $\frac{1}{3}<p\leq1$, where $\rho_{\frac{1}{3}}^{BSA}=\frac{1}{3}|\psi^+\rangle\langle\psi^+|+\frac{2}{3}\cdot\frac{I}{4}$ \cite{Englert00,Wellens01}. Mixing $|\psi^+\rangle$ with $|\psi^-\rangle=\frac{1}{\sqrt{2}}(|00\rangle-|11\rangle)$, $\eta_q=q|\psi^-\rangle\langle\psi^-|+(1-q)|\psi^+\rangle\langle\psi^+|$.
Interestingly, $\eta_{\frac{1}{2}}$ is separable, but $\eta_{\frac{3}{4}}$ belongs to the family of $|\psi^-\rangle$ instead of the family of $|\psi^+\rangle$. Figure 2(b) illustrates the schematic picture.

The set of Werner states forms part of the family of entangled states that has the Bell state at its core. It has often been used to successfully verify some results \cite{Horodecki09,Guhne09}. However, there exist states in the family associated with the Bell state which are not Werner states. This incompleteness, together with the low dimension, may explain why some results cannot be applied to the Werner states (e.g. the Horodeci states \cite{Horodecki99}).


\noindent

\textit{Conclusions and discussions.---}
We have showed a framework where entangled states play the role
of witnesses. We answer the open question of what the remainder of the Lewenstein-Sanpera decomposition of a density matrix. We gave an unambiguous and non-numeric quantification of entanglement, and gave a family-structure classification and a completely new structure description of entangled states. We argue that we cannot simply tell which one is more entangled than another one between entangled states in different families. Our results indicate that we can focus only on the study of a typical family of entangled states when we investigate entangled states.
Our framework assembles many seemingly different findings with simple arguments that do not require lengthy calculations.


Here we mainly consider the case of discrete bipartite systems on the finite-dimensional Hilbert space, but we can also generalize our results to continuous-variable systems, multipartite systems and infinite-dimensional Hilbert space because they all have a common mathematical foundation and physical interpretation. However, entanglement in continuous-variable systems like harmonic oscillators or light modes is significantly different \cite{ContinuousVariables} from the case of the discrete systems. We have not discussed these systems here. Recently, Demianowicz and Augusiak \cite{Demianowicz} showed a method for constructing optimal entangled states (genuinely entangled states) in a multiparty scenario by finding a genuinely entangled subspace. The result also shows evidence that our results can be generalized into the multiparty scenario. However, the complexity of the separability problem increases substantially when we study multipartite systems \cite{Gerke16}, and the structure of multipartite entanglement is much richer than the bipartite entanglement \cite{Gerke16,Wang14,Rossi13,Huber13,Sperling13,Shahandeh14}.

\begin{acknowledgments}
We would like to especially thank Christopher Perry, Guang Ping He, and Reevu Maity for revising the original manuscript, and Maciej Lewenstein very helpful discussions. We would also like to thank the anonymous for their helpful comments and suggestions. We are grateful to Christopher Perry, Dan Brown, Jonathan Oppenheim, Fernando Brand\~{a}o, Guang Ping He, Marco
Piani, Zhihao Ma, Yu Guo, Shao-Ming Fei, Reevu Maity, Xiao Yuan, Leong-Chuan Kwek, Otfried G\"{u}hne and Vlatko Vedral for helpful discussions and suggestions. We thank Dong-Yang Long at Sun Yat-sen University where this
work was started while Wang was a doctoral candidate, Simone Severini for his kind hospitality at University College London, Leong-Chuan Kwek for his kind hospitality at National University of Singapore and Vlatko Vedral for his kind hospitality at University of Oxford where part of this work was (is) carried out while Wang was (is) an academic visitor. This work is supported by the National Natural
Science Foundation of China under Grants No. 61672007 and No. 61272013.
\end{acknowledgments}



\begin{thebibliography}{99}

\bibitem {Horodecki09} R. Horodecki,
P. Horodecki, M. Horodecki and K. Horodecki, {\it Rev. Mod. Phys. }
{\bf81} 865 (2009).

\bibitem {Guhne09} O. G\"{u}hne and G. T\'{o}th, {\it Phys. Rep.}{\bf 474} 1 (2009).

\bibitem{Plenio07} Martin B. Plenio and Shashank Virmani, {\it Quantum Information and Computation }
{\bf7} 1 (2007).



\bibitem {Lewenstein98} Maciej Lewenstein, Anna Sanpera, {\it Phys. Rev. Lett.} {\bf 80} 2261 (1998).

\bibitem{Sanpera98} A. Sanpera, R. Tarrach, and Guifr\'{e} Vidal, {\it Phys. Rev. A} {\bf 58} 826 (1998).


\bibitem{Kraus00} B. Kraus, J. I. Cirac, S. Karnas, and M. Lewenstein, {\it Phys. Rev. A} {\bf 61}
062302 (2000).
\bibitem{Horodecki00} Pawel Horodecki, Maciej Lewenstein, Guifr¨¦ Vidal, Ignacio Cirac, {\it Phys. Rev. A} {\bf 62}
032310 (2000).
\bibitem {Lewenstein00} M. Lewenstein, B. Kraus, J. I. Cirac and P. Horodecki, {\it Phys. Rev. A} {\bf 62}
052310 (2000).
\bibitem {Karnas01} S. Karnas and M. Lewenstein, {\it J. Phys. A: Math. Gen.} {\bf 34} 6919 (2001).

\bibitem {Thiang09} Guo Chuan Thiang, Philippe Raynal, and Berthold-Georg Englert, {\it Phys. Rev. A} {\bf 80} 052313 (2009).

\bibitem {Thiang10} Guo Chuan Thiang, {\it Phys. Rev. A} {\bf 82} 012332 (2010).
\bibitem {Quesada14} R. Quesada, A. Sanpera, {\it Phys. Rev. A} {\bf 89} 052319 (2014).
\bibitem {Lewenstein16} M. Lewenstein, R. Augusiak, D. Chru\'{s}ci\'{n}ski, S. Rana, and J. Samsonowicz {\it Phys. Rev. A} {\bf 93} 042335 (2016).



\bibitem {Chruscinski14} D. Chru\'{s}ci\'{n}ski, and G. Sarbicki (2014), {\it  J. Phys. A: Math. Theor. }{\bf 47}
483001 (2014).

\bibitem {Augusiak11} R. Augusiak, J. Bae, {\L}. Czekaj and M.
Lewenstein, {\it J. Phys. A} {\bf 44} 185308 (2011).
\bibitem {Edwards65} R.E. Edwards, Functional analysis, theory and application (Holt, Rinehart and Winston. New York, 1965).

\bibitem {M.Horodecki96} M. Horodecki, P. Horodecki and R. Horodecki, {\it Phys. Lett. A} {\bf223}
1 (1996).
\bibitem {Terhal00} B. M. Terhal, {\it Phys. Lett. A }{\bf 271}
319 (2000).


\bibitem {Doherty04} A. C. Doherty, P. A. Parrilo, and F. M. Spedalieri, {\it Phys. Rev. A }{\bf 69}
022308 (2004).
\bibitem {Wu06} YuChun Wu, YongJian Han, and GuangCan Guo, {\it Phys. Lett. A }{\bf 356}
402 (2006).
\bibitem {Wu07} Y.-C. Wu and G.-C. Guo, {\it Phys. Rev. A} {\bf 75}
052333 (2007).
\bibitem {Sperling09} J. Sperling and W. Vogel, {\it Phys. Rev. A }{\bf 79}
022318 (2009).
\bibitem {Chruscinski09} D. Chru\'{s}ci\'{n}ski, J. Pytel
and G. Sarbicki, {\it Phys. Rev. A} {\bf 80} 062314 (2009).
\bibitem {Hou10a} J. Hou and Y. Guo, {\it Phys. Rev. A }{\bf 82}
052301 (2010).
\bibitem {Chruscinski10} D. Chru\'{s}ci\'{n}ski and J. Pytel, {\it Phys. Rev. A} {\bf 82} 052310 (2010).
\bibitem{Wang11} B.-H. Wang and D.-Y. Long, {\it Phys. Rev. A }{\bf 84}
014303 (2011).


\bibitem {Gurvits04}  L. Gurvits, {\it J. Comput. Syst. Sci.}, {\bf69} 448 (2004).

\bibitem{Lewenstein01} M. Lewenstein, B. Kraus, P. Horodecki, and J. I. Cirac, {\it Phys. Rev. A} {\bf 63}
044304 (2001).



\bibitem{Brub2002} D. Bru{\ss},J. I. Cirac, P. Horodecki, F. Hulpke, B. Kraus, M. Lewenstein, and A.
Sanpera, {\it Journal of Modern Optics} {\bf 49} 1399 (2002). arxiv: quant-ph/0110081.

\bibitem {Milne15} A. Milne, D. Jennings, and T. Rudolph, {\it Phys. Rev. A }{\bf 92} 012311 (2015).

\bibitem {Bender98} C. M. Bender and S. Boettcher {\it Phys. Rev. Lett.} {\bf 80} 5243 (1998).

\bibitem {Moiseyev11} N. Moiseyev, \emph{Non-Hermitian Quantum Mechanics} (Cambridge University Press, Cambridge, England, 2011)

\bibitem {Pati15} A. K. Pati, U. Singh, and U. Sinha, {\it Phys. Rev. A }{\bf 92} 052120 (2015).



\bibitem{Sarbicki07} G. Sarbicki, General theory of detection and optimality. arXiv:0905.0778V2 (2007).

\bibitem{Piani16} Marco Piani, Marco Cianciaruso, Thomas R. Bromley, Carmine Napoli, Nathaniel Johnston, and Gerardo Adesso, {\it Phys. Rev. A} {\bf 93}
042107 (2016).

\bibitem {Napoli16} Carmine Napoli, Thomas R. Bromley, Marco Cianciaruso, Marco Piani, Nathaniel Johnston, and Gerardo Adesso, {\it Phys. Rev. Lett.} {\bf 116} 150502 (2016).

\bibitem{SuperCW} The properties and characterization of coherent states as high-level coherence witnesesses are under preparation.


\bibitem{Shahandeh17} F. Shahandeh, M. Ringbauer, J.C. Loredo, and T. C. Ralph, {\it Phys. Rev. Lett.} {\bf 118} 110502 (2017).

\bibitem{Terhal2000} B. M. Terhal, and P. Horodecki, {\it Phys. Rev. A} {\bf 61} 040301(R) (2000).

\bibitem{Sanpera01} A. Sanpera, D. Bru{\ss}, and M. Lewenstein, {\it Phys. Rev. A} {\bf 63} 050301(R) (2001).

\bibitem{Bruns16} D. Bruns, J. Sperling, and S. Scheel, {\it Phys. Rev. A} {\bf 93}
032132 (2016).

\bibitem{Vedral97} V. Vedral, M. B. Plenio, M. A. Rippin, and P. L. Knight,  {\it Phys. Rev.  Lett. }{\bf 78}
2275 (1997).



\bibitem {Brandao05}F. G. S. L. Brand\~{a}o, {\it Phys. Rev. A} {\bf 72}
022310 (2005).

\bibitem{Huang14} Yichen Huang,  {\it New J. Phys. }{\bf 16}
033027 (2014).

\bibitem{Bennett96} C. H. Bennett, D. P. DiVincenzo, J. A. Smolin, and W. K. Wootters, {\it Phys. Rev. A} {\bf 54} 3824 (1996).
\bibitem{Hill97} S. Hill and W. K. Wootters {\it Phys. Rev.  Lett. }{\bf 78}
5022 (1997).
\bibitem{Wootters98} W. K. Wootters {\it Phys. Rev.  Lett. }{\bf 80}
2245 (1998).
\bibitem{Zyczkowski98} K. \.{Z}yczkowski, P. Horodecki, A. Sanpera, and M. Lewenstein, {\it Phys. Rev. A} {\bf 58} 883 (1998).
\bibitem{Vidal02} G. Vidal, and R. F. Werner {\it Phys. Rev. A} {\bf 65} 032314 (2002).


\bibitem {Shahandeh14}F. Shahandeh, J. Sperling, and W. Vogel, {\it Phys. Rev. Lett.}
{\bf 113} 260502 (2014).
\bibitem {Regula16}B. Regula, and G. Adesso, {\it Phys. Rev. Lett.}
{\bf 116} 070504 (2016).



\bibitem{Leinaas07} J. M. Leinaas, J. Myrheim and E. Ovrum, {\it Phys. Rev. A} {\bf 76}
034304 (2007).

\bibitem {Horodecki97} P. Horodecki, {\it Phys. Lett. A} {\bf 232} 333 (1997)
\bibitem {Wallach02}N. R. Wallach. {\it Contemp. Math.} {\bf 305} 291 (2002).

\bibitem {Parthasarathy04}K. R. Parthasarathy. {\it Proc. Indian Acad. Sci.
(Math.Sci)} {\bf 114} 365 (2004).

\bibitem {Cubitt08}T. Cubitt, A. Montanaro, and A. Winter. {\it J. Math.
Phys.} {\bf 49} 022107 (2008).

\bibitem {Walgate08}J. Walgate and A. Scott. {\it J. Phys. A: Math. Theor.}
{\bf 41} 375305 (2008).






\bibitem {Werner89}  R. F. Werner, {\it Phys. Rev. A} {\bf 40}
4277 (1989).
\bibitem {Englert00} B.-G. Englert and N. Metwally, {\it J. Mod. Opt.} {\bf 47} 2221 (2000).
\bibitem {Wellens01} T. Wellens and M. Ku\'{s}, {\it Phys. Rev. A} {\bf 64} 052302 (2001).

\bibitem {Horodecki99} P. Horodecki, M. Horodecki, and R. Horodecki, {\it Phys. Rev. Lett.} {\bf 82}
1056 (1999).



\bibitem {ContinuousVariables} For Gaussian states in systems with continuous variables by the Hahn-Banach theorem, the sets must be closed and convex, but not necessarily bounded \cite{Brub2002}.

\bibitem{Demianowicz} M. Demianowicz and R. Augusiak, arXiv:1712.08916 (2017).
\bibitem {Gerke16} S. Gerke, J. Sperling, W. Vogel, Y. Cai, J. Roslund, N. Treps, and C. Fabre {\it Phys. Rev. Lett.} {\bf 117} 110502 (2016).
\bibitem {Wang14} B.-H. Wang, H.-R. Xu, and S. Severini {\it Phys. Rev. A}
{\bf 90} 022312 (2014).
\bibitem {Sperling13}J. Sperling and W. Vogel, {\it Phys. Rev. Lett.}
{\bf 111} 110503 (2013).

\bibitem {Rossi13} M. Rossi, M. Huber, D. Bru{\ss} and C. Macchiavello, {\it New J. Phys.}
{\bf 15} 113022 (2013).
\bibitem {Huber13} M. Huber and Julio I. de Vicente {\it Phys. Rev. Lett.}
{\bf 110} 030501 (2013).





\end{thebibliography}
\end{document}